\documentclass[twocolumn]{aastex631}
\pdfoutput=1
\usepackage{mathtools}
\usepackage{appendix}
\usepackage{bm}
\usepackage{natbib}
\usepackage{orcidlink}
\usepackage{newtxtext,newtxmath}
\usepackage[T1]{fontenc}
\usepackage{ae,aecompl}
\usepackage{chemformula}
\usepackage{booktabs}
\usepackage{xcolor}
\usepackage{gensymb}
\usepackage{amsmath}
\usepackage{threeparttable,booktabs}

\newcommand{\MJup}{M_\mathrm{Jup}}
\newcommand{\RJup}{R_\mathrm{Jup}}

\shorttitle{ZTF J1239+8347}
\shortauthors{Whitebook et al.}
\submitjournal{ApJ Letters}

\begin{document}

\title{A Mass Transferring Brown Dwarf Binary on a 57 Minute Orbit}

\author[0000-0002-6836-181X]{Samuel Whitebook}
\affil{Division of Physics, Mathematics, and Astronomy, California Institute of Technology, Pasadena, CA 91125, USA}

\author[0000-0003-4189-9668]{Antonio C. Rodriguez}
\affil{Center for Astrophysics, Harvard \& Smithsonian, 60 Garden Street, Cambridge, MA 02138, USA}
\affil{Division of Physics, Mathematics, and Astronomy, California Institute of Technology, Pasadena, CA 91125, USA}

\author[0000-0002-7226-836X]{Kevin Burdge}
\affil{Department of Physics, Massachusetts Institute of Technology, Cambridge, MA 02139, USA}

\author[0000-0003-2630-8073]{Thomas Prince}
\affil{Division of Physics, Mathematics, and Astronomy, California Institute of Technology, Pasadena, CA 91125, USA}

\author[0000-0002-8895-4735]{Dimitri Mawet}
\affil{Division of Physics, Mathematics, and Astronomy, California Institute of Technology, Pasadena, CA 91125, USA}

\author[0000-0003-4725-4481]{Sam Rose}
\affil{Division of Physics, Mathematics, and Astronomy, California Institute of Technology, Pasadena, CA 91125, USA}

\author[0000-0002-4717-5102]{Pablo Rodr\'{\i}guez-Gil}
\affil{Instituto de Astrofísica de Canarias, E-38205 La Laguna, Tenerife, Spain}
\affil{Departamento de Astrofísica, Universidad de La Laguna, E-38206 La Laguna, Tenerife, Spain}

\author{Anica Ancheta}
\affil{Division of Physics, Mathematics, and Astronomy, California Institute of Technology, Pasadena, CA 91125, USA}

\author{Ariana Pearson}
\affil{Department of Physics and Astronomy, University of Waterloo, 200 University Avenue West, Waterloo, Ontario N2L 3G1, Canada}

\author[0000-0002-8623-8268]{Sage Santomenna}
\affil{Department of Physics and Astronomy, Pomona College, 333 N.\ College Way, Claremont, CA 91711, USA}

\author[0000-0002-5812-3236]{Aaron Householder}
\altaffiliation{NSF Graduate Research Fellow}
\affiliation{Department of Earth, Atmospheric and Planetary Sciences, Massachusetts Institute of Technology, Cambridge, MA 02139, USA}
\affil{Kavli Institute for Astrophysics and Space Research, Massachusetts Institute of Technology, Cambridge, MA 02139, USA}

\author[0000-0002-6618-1137]{Jerry W. Xuan}
\altaffiliation{51 Pegasi b Fellow}
\affil{Department of Earth, Planetary, and Space Sciences, University of California, Los Angeles, CA 90095, USA}
\affil{Division of Physics, Mathematics, and Astronomy, California Institute of Technology, Pasadena, CA 91125, USA}

\correspondingauthor{Samuel Whitebook}
\email{sewhitebook@astro.caltech.edu}

\begin{abstract}
Mass transfer in stellar binaries has been well studied in most stellar mass ranges, with the notable exception of ultracool stars and substellar brown dwarfs. We report the discovery of ZTF J1239+8347 with the Zwicky Transient Facility (ZTF), a brown dwarf binary currently undergoing stable mass transfer with an orbital period of 57.41 minutes. Optical time-series photometry reveals an extremely high amplitude ($> 2$ magnitude peak-to-trough) variability at short wavelengths indicative of an orbiting hot spot slightly buried inside the atmosphere of the accretor. We use parallax measurements from \textit{Gaia} along with optical and near infrared spectra to infer an accretion temperature of $T_\mathrm{eff} = 8904 \pm 54$ K, an atmospheric temperature of the accretor of $T_\mathrm{atmo} \approx 1500$ K, and a slightly inflated accretor radius of  $R_{\rm acc} = 1.20^{+0.15}_{-0.11} \, \RJup$. ZTF J1239+8347 is a direct impact accretor, typically only seen in double degenerate white dwarf binaries, which are approximately a million times denser than the components in ZTF J1239+8347. The existence of an accreting brown dwarf binary suggests that angular momentum loss can be strong enough to make ultracool binaries interact in a Hubble time. The observed faintness ($\sim 20$ mag) and relative proximity ($\approx 300$ pc) of ZTF J1239+8347 suggests that many similar systems are likely to be found by the upcoming Rubin Observatory Legacy Survey of Space and Time (LSST).
\end{abstract}

\section{Introduction} \label{introduction}

Binary interaction is responsible for many types of exotic stars and unique phenomena. Thousands of main sequence (MS) binaries have been discovered over the years in several photometric surveys such as the All Sky Automated Survey (ASAS, \citealp{Pojmanski_ASAS_1, Pojmanski_ASAS_Variables}), Sloan Digital Sky Survey (SDSS, \citealp{York_SDSS_Technical}), and Catalina Real Time Sky Survey (CRTS, \citealp{djorgovski2011catalinarealtimetransientsurvey, Drake_CRTS_Variables}). Binaries consisting of one or more degenerate objects, such as X-ray binaries consisting of a neutron star or black hole and a MS star \citep[e.g.][]{White_XBs}, create many complex phenomena and rich periodic signatures when their orbits decay, one component overflows its Roche lobe, and mass transfer begins. In this paper we present the discovery of a mass transferring brown dwarf binary.

Brown dwarfs (BDs) bridge the gap between stars and planets. These substellar objects lie below the minimum mass required to fuse hydrogen \citep{Burrows_BDs_Overview} and possess fully convective interiors and partially-degenerate cores \citep{Burrows_BDs_Theory, chabrier1999physicsbrowndwarfs}. Without a central engine, BDs slowly radiate away their formation energy and fade through the spectral types M, L, T, and Y \citep{Kirkpatrick+Reid+Liebert+etal_1999, Kirkpatrick+Gelino+Cushing+etal_2012}. Since these objects lack distinct evolutionary phases, after $\sim 1$ Gyr BDs contract slowly \citep{baraffe1997atmospheremodelslowmass}, and binary BDs lack an evolutionary angular momentum loss mechanism to bring them in from their formation orbits to a separation at which they can interact. However, many BDs form in the presence of a larger MS star or a third brown dwarf, with several known cases of BD binaries existing in triples, even amongst the nearest BDs (e.g. \citealp{McCaughrean_EpsIndi, Triaud_BDTriple, Whitebook_GJ229B, Xuan_Gl229B}). Of note are several BD binaries within each others' Hill spheres \citep[e.g.][]{Stassun_BD_Binary, Xuan_Gl229B}, including the T dwarf binary Gliese 229 B. The BD components of Gliese 229 B themselves have an eccentric orbit around an M dwarf host, Gliese 229 A. The existence of this system at such a short period requires mechanisms that dissipate significant energy in substellar binaries \citep{Xuan_Gl229B}, bringing them close enough to reach interaction separations. It is not yet clear what those mechanisms are, and at least for Gliese 229 A-Bab, the orbital mutual inclinations are below the threshold for Kozai-Lidov oscillations \citep{Thompson2025}. Another similar object to ZTF J1239+8347 is the almost accreting system ZTF J2020+5033 consisting of a high mass BD transiting a low mass MS star almost at their Roche limit \citep{El_Badry_BD}. ZTF J2020+5033 is closer than the vast majority of other known transiting BDs, at only $\approx 140$ pc, which provides further evidence that BDs reaching short orbits are not very rare. This system is likely a progenitor of the accreting brown dwarf binaries (hereafter aBDBs), and Roche lobe overflow is predicted to occur within only $\lesssim 1$ Gyr. There are several other relatively tight ultracompact binaries known \citep[e.g.][]{Jackman_UCB, Hsu_short_BDB} as well as cataclysmic variable systems with BD donors \citep[e.g.][]{Neustroev_BD_CV, Galiullin_BD_CV}.

It is theorized that low-mass MS stars in binaries can not maintain stable mass transfer \citep{Rucinski_Short_Period_Limit, Jiang_Short_Period_Limit}. Being semi-degenerate, BDs possess very shallow mass-radius relations, and for high mass BDs ($M \gtrsim 60 \MJup$) the adiabatic response to mass loss is to contract, which would serve to keep the systems stable under mass transfer.

A physical analog of the semi-degenerate aBDBs are interacting double white dwarf (WD) binaries, which consist of two fully degenerate components. These systems can maintain accretion disks at longer periods, but at the shortest periods they become direct impact accretors \citep{Marsh_Direct_Impact, Marsh_DWD, Deloye_AMCVN} as we predict all aBDBs to be (section \ref{sec:results}). These systems maintain stable mass transfer through angular momentum feedback from tidal forces coupling accretor spin to the orbit until they lose too much momentum through gravitational radiation and magnetic braking and merge \citep{Marsh_DWD}.

The Zwicky Transient Facility \citep[ZTF;][]{Graham_ZTF, Bellm_ZTF, Masci_ZTF} has undertaken a comprehensive search for periodic variable stars in the ZTF Variability survey (ZVAR) (Whitebook et al., in prep.). Within the ZVAR survey we undertook a targeted search for aBDBs. In this search we identified ZTF J1239+8347; which exhibits large amplitude optical variability on a 57 minute period, but is extremely dim ($m_g < 22.5$) at minimum, despite its relatively close distance of $339^{+42}_{-31}$ pc \citep{Bailer-Jones_Distance}. Despite this large optical variability, ZTF J1239+8347 is incompatible with known types of compact object binaries including cataclysmic variable stars (CVs) and black widow binaries (BWs) (see Section~\ref{subsec:compact_binaries}).

\section{Search Criteria and Discovery} \label{sec:discovery}
ZTF J1239+8347 was found in a search of the ZVAR periodic variables dataset. ZVAR utilizes the Fast Periodicity Weighting (FPW) period finding algorithm, which fits piecewise constant phase bins to test frequencies \citep{finkbeiner2025_FPW}, to detect periodicity in ZTF light curves for 1.5 billion stars with $m_g >  22$. We utilized photometry and astrometry from the \textit{Gaia} space satellite \citep{GaiaMission, GaiaDR3, Riello_GaiaDR3} to define the parameter space of likely aBDB candidates, based on the hypothesis that these systems should appear similar to BWs consisting of a low-mass star and a neutron star \citep[T. Marsh, private communication;][]{Burdge_BW}. We utilize constraints and predictions from theoretical dynamics of aBDB systems (Whitebook et al., in prep.) to construct our search criteria, particularly our period range and light-curve morphology. We restricted our search to sources with \textit{Gaia} photometry lying below the MS with $BP-RP> 0.0$ and $\Delta M_G \geq 2$. Additionally we limited the search to a period range of $50$ minutes to $3.5$ hours to include the theoretical range of aBDBs. We hand picked candidates that had flat bottom phase folded light curves with positive excursions. Of these, only ZTF J1239+8347 has a well constrained parallax, which was necessary to help rule out the source as a BW binary (see Section~\ref{sec:discussion}). ZTF J1239+8347 has previously been identified as a WD and cataclysmic variable candidate \citep{Fusillo_WD_Imposter1, Fusillo_WD_Imposter2, Ren_CVs, Steen_CV_Imposter}.

\section{Photometry and Spectra} \label{sec:followup}
ZTF J1239+8347 shows starkly different features in the optical versus the infrared. In the optical and UV the system is extremely variable from accretion luminosity, while in near and mid infrared the accretion luminosity is dim enough that the thermal emission from the BD atmospheres contributes significantly to the spectrum. In addition to the optical and infrared data, we observed ZTF J1239+8347 with the Neil Gehrels Swift (Swift) Observatory \citep{Burrows_SWIFT}. No X-ray emission was detected in the $0.3 - 10$ keV band; assuming $N_\mathrm{H} = 10^{20}\,\mathrm{cm}^{-2}$ and a power-law index of 2, we place an upper limit on the X-ray flux of $\sim 10^{-13}\,\mathrm{erg\,s^{-1}\,cm^{-2}}$.

\subsection{Optical} \label{sec:optical}
\begin{figure}[t!]
    \centering
    \includegraphics[width=1\linewidth]{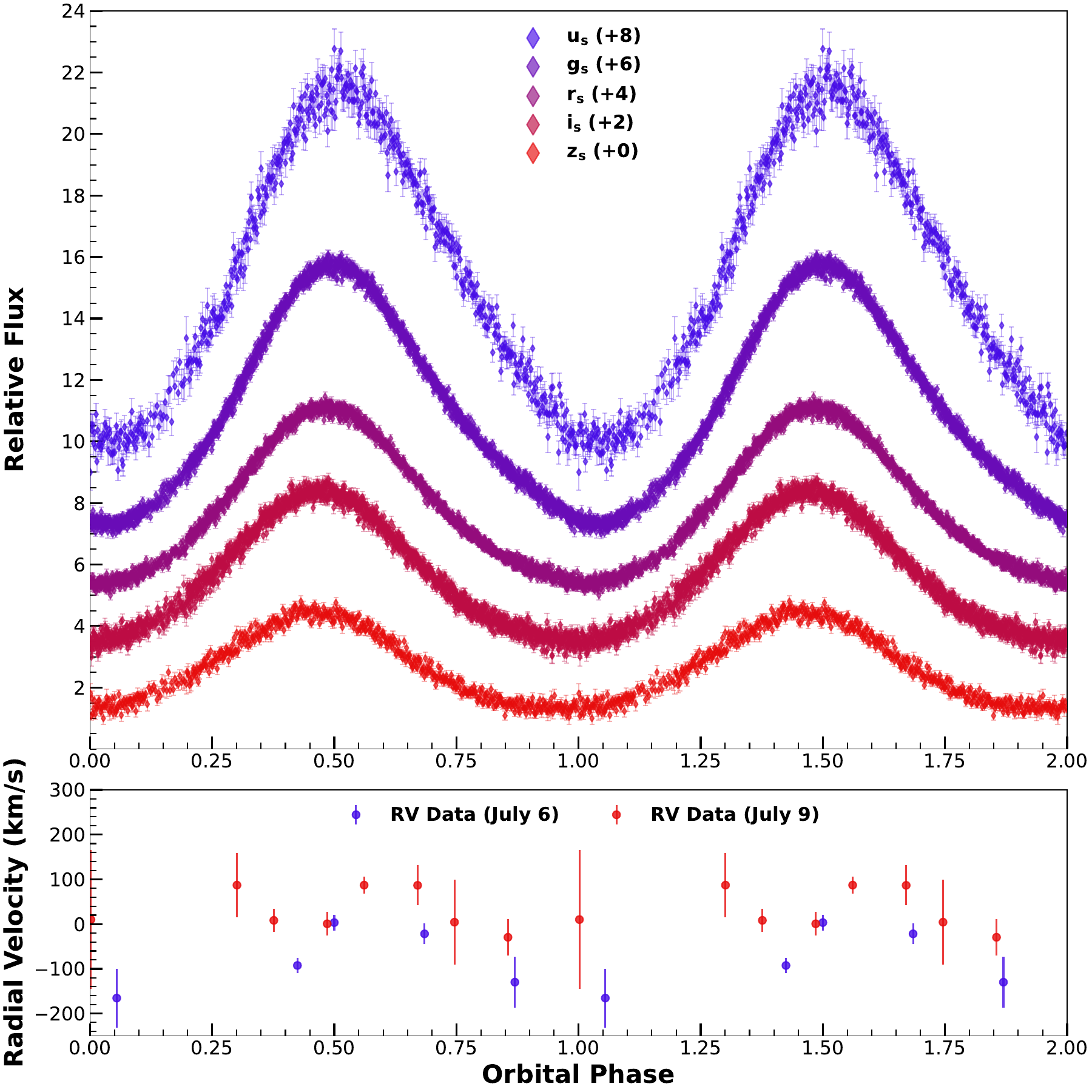}
    \caption{Top: GTC/HiPERCAM  light curves of ZTF J1239 in the \textit{u\textsubscript{s}}, \textit{g\textsubscript{s}}, \textit{r\textsubscript{s}}, \textit{i\textsubscript{s}}, and \textit{z\textsubscript{s}} bands. Each light curve is normalized with respect to its minimum flux level and offset for visibility. 
    Bottom: Radial velocity (RV) data from Keck/LRIS (see \ref{sec:optical}). Light curves are phased such that the maximum of the $u_s$ curve is at $\phi = 0.5$. RV epochs are phased such that the epoch of the brightest LRIS observation is coincident with the phase of the maximum of the $g_s$ light curve.}
    \label{fig:photometry}
\end{figure}

ZTF J1239+8347 was discovered due to its extreme variability in g and r band ZTF light curves. At minimum brightness the source reaches the detection limit of ZTF ($m \sim 22.5$) in all bands. We obtained follow up photometry with the HiPERCAM camera \citep{Dhillon_HIPERCAM} on the Gran Telescopio Canarias (GTC) on 2025 June 27, simultaneously observing in five bands (see Figure \ref{fig:photometry}). Exposure times were 3.8\,s in \textit{u\textsubscript{s}} and 1.0\,s in \textit{g\textsubscript{s}}, \textit{r\textsubscript{s}}, \textit{i\textsubscript{s}}, and \textit{z\textsubscript{s}}, with a total observing duration of 2 hours. The data were reduced using the standard HiPERCAM reduction pipeline \citep{Dhillon_HIPERCAM}.

\begin{figure*}[th!]\centering
    \includegraphics[width =\linewidth]{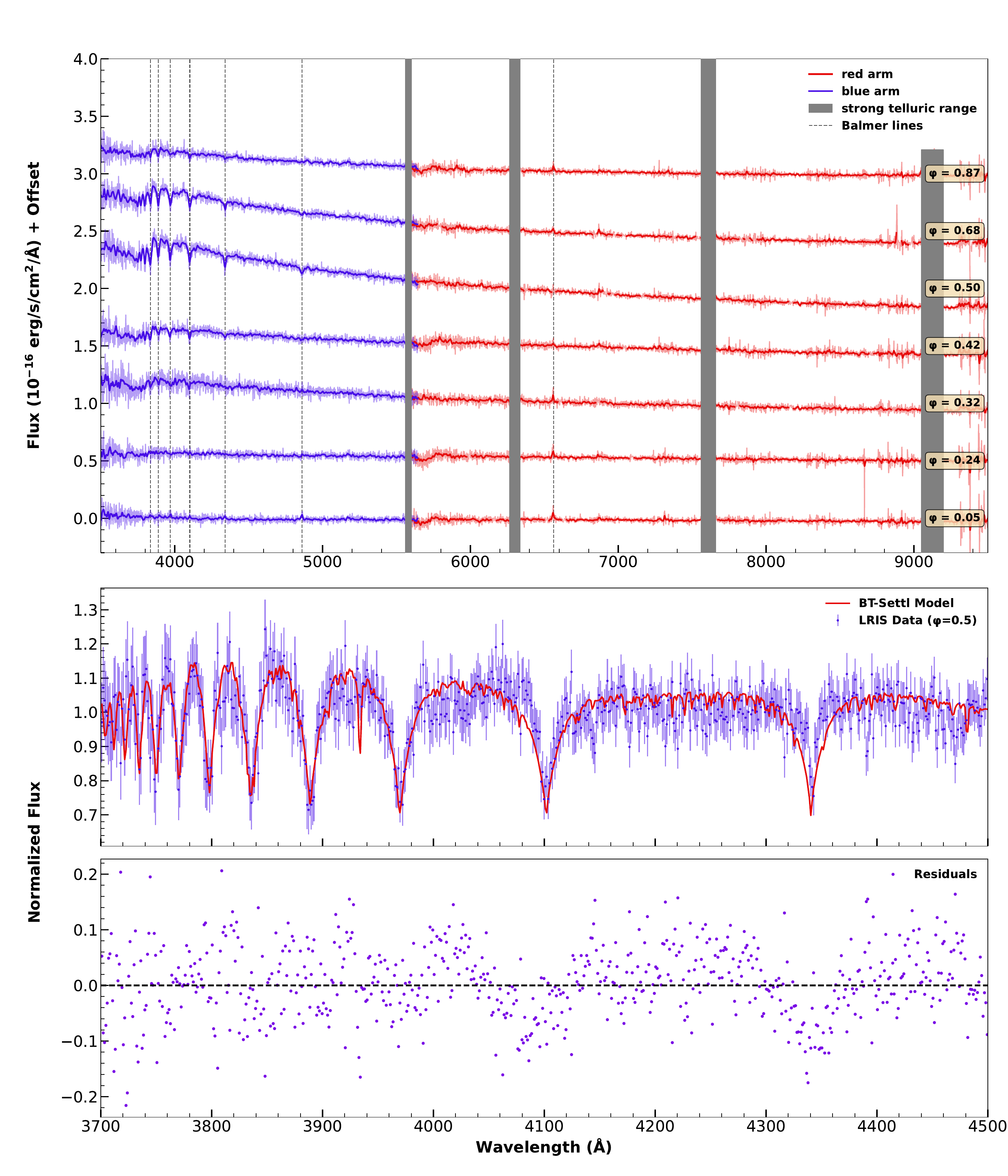}
    \caption{Top: Phase resolved LRIS spectra of ZTF J1239+8347. Full resolution spectra are shown behind spectra smoothed to $R \sim 800$ for clarity. The spectra are each offset by 0.25 from the next for clarity and phased such that the brightest spectrum is at $\phi = 0.5$ for consistency with optical photometry. Hydrogen absorption features are prominently visible during bright times, and disappear to a mostly featureless flat spectrum with H-$\alpha$ emission at minimum suggesting a shock. Middle: The normalized spectrum of the brightest LRIS epoch with the normalized \texttt{BT-Settl} model cut to the Balmer region used for RV fitting. Bottom: Residuals of the \texttt{BT-Settl} model subtracted from the normalized LRIS spectrum.}
    \label{fig:lris}
\end{figure*}

Phase-resolved optical spectra were obtained with the Low-Resolution Imaging Spectrograph (LRIS, \citealp{Oke_LRIS, Rockosi_LRIS}) on the Keck I telescope on 2024 July 6 and 2024 July 9 in the 600/4000 disperser mode ($R \sim 3200$) as part of a survey of polar CV candidates. LRIS data were reduced using the \texttt{LPIPE} reduction package \citep{Pearly_Lpipe}. The LRIS spectra from 2024 July 6 are plotted separated by phase in figure \ref{fig:lris}. The spectrum manifests as a mostly featureless blackbody that flattens towards minimum brightness. At maximum brightness hydrogen absorption features appear prominently blueward of 4400 \AA. Conversely, when the system is at minimum brightness, H-$\alpha$ appears as a broad emission feature that disappears as the blackbody comes into view. We fit the brightest spectrum with a Maxwell-Boltzmann distribution with a $\chi^2$ grid fit of the temperature, $T_\mathrm{eff}$, and marginalize over an arbitrary scale parameter for each temperature. We bootstrap resample the spectrum, drawing with replacement for $100$ iterations and take the mean as the best fit temperature and the standard deviation as the error. We derive a $T_\mathrm{eff} = 8904 \pm 54$ K.

We compute the observed radial velocity of the hydrogen absorption features by fitting each LRIS blue arm spectrum from both nights with a \texttt{BT-Settl} model spectrum of an A type star, which provides a close match to the observed features \citep{Allard_BT-Settl}. We adopt a model temperature of $8800$ K to approximate the observed accretion temperature, assuming solar metallicity and $\log(g) = 4.5$ (figure \ref{fig:lris}). We restrict our fitting to the range of $3700 < \lambda < 4500 \,$ \AA \, to best capture the most prominent Balmer features. We degrade the model spectrum to the LRIS resolution with a Gaussian filter and divide out a $7$ degree polynomial from the model and each spectrum. Since the absorption depths change significantly between epochs, we scale the model flux absorption to best match the absorption depths observed in the data as $f_\mathrm{scale} = 1 - s(1 - f)$ where $s$ is a fit parameter between 0 and 1. We fit each epoch for an RV by interpolating the model over a grid of redshifts. We joint fit for the per epoch RV and model absorption depth and pick the values which minimize $\chi^2$. Uncertainties on the individual RVs are derived from $\Delta \chi^2_{\mathrm RV} = 1$ bounds. We find that the residuals of the template fits are consistently Gaussian with $|\mu| < 0.1$ and $0.9 < \sigma < 1.1$ indicating that template mismatch does not impose a significant systematic in the RV fits. Individual epoch RVs are shown in figure \ref{fig:photometry}.

\subsection{Infrared}
ZTF J1239+8347 is too dim in the near infrared (NIR) to be detected by any past JHK sky surveys, including the Two Micron All Sky Survey (2MASS). We obtained \textit{K} band photometry of ZTF J1239+8347 on 2025 July 8 with the Palomar Wide field InfraRed Camera (WIRC, \citealp{Wilson_WIRC}). We observed with a 3 second exposure, 10 coadd, 1x9 dither pattern for 1 hour and derive an AB \textit{K} magnitude of $21.2$ $\pm 1.2$. In addition to this, we obtained J band photometry with WIRC on 2025 July 29 with 10 second exposures, 1 coadd, 1x9 dither pattern for 1 hour from which we derive an AB \textit{J} magnitude of $20.3$ $\pm 0.8$. NIR photometric measurements were taken over a single orbital period and represent orbital averages. Data was stacked over the full observation time and extracted with the SExtractor pipeline \citep{Bertin_SExtractor}. Epoch WIRC data is too low SNR to significantly detect photometric variability within an orbital period, which is contained within uncertainties.

\begin{figure*}[th!]\centering
    \includegraphics[width =\linewidth]{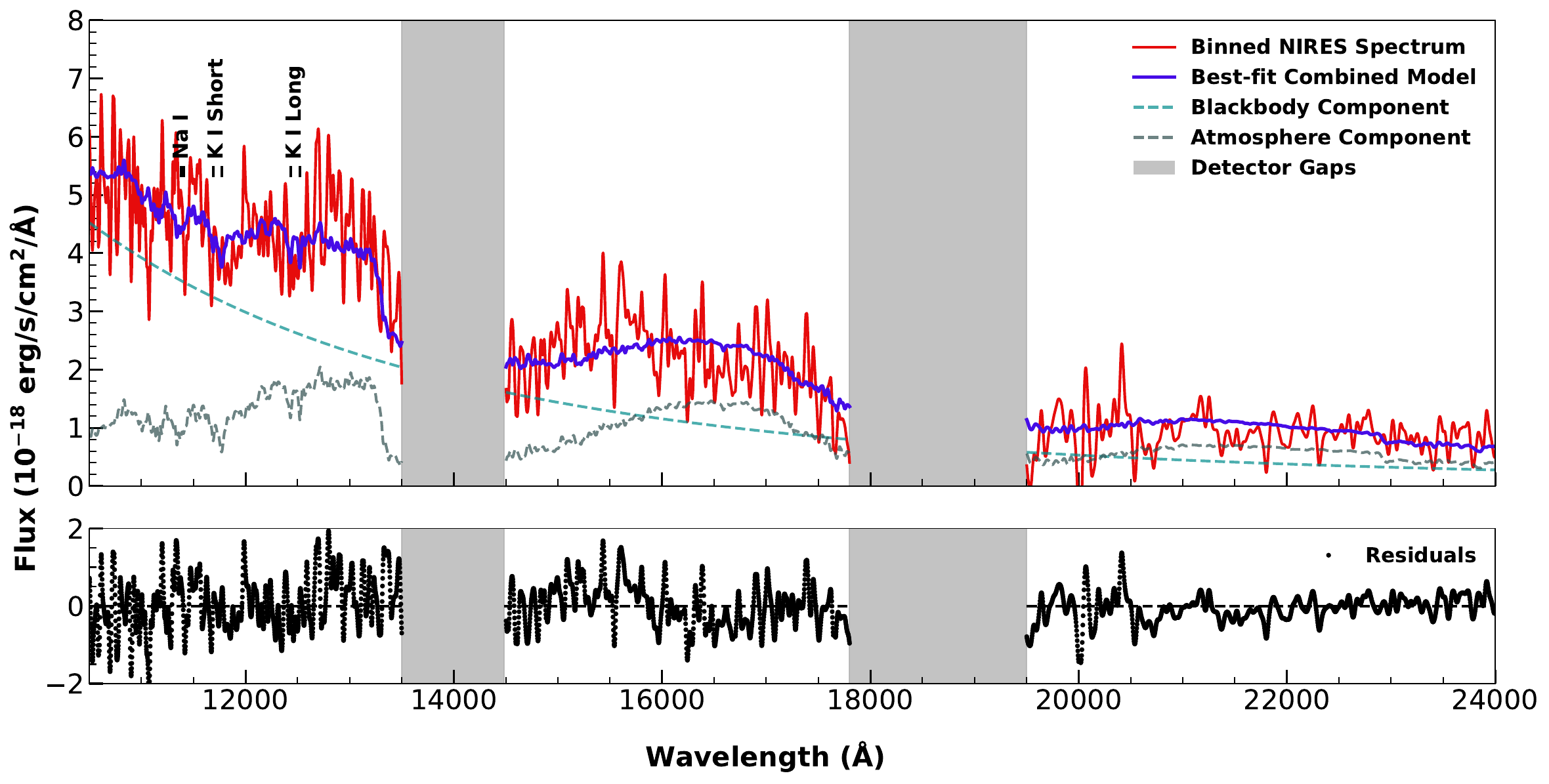}
    \caption{NIRES spectrum of ZTF J1239+8347 with the best fit scaled blackbody plus single BD atmosphere model overlaid. The shown spectrum is a sum of data over a full orbital period and is smoothed for clarity. The BD model atmosphere chosen is the one with the atmospheric temperature that best fits the data and provides a physically plausible BD radius of $R_{\rm BD} = 1.20^{+0.15}_{-0.11} \, \RJup$}; $T_\mathrm{atmo} \approx 1500$ K. Na and K doublet features are marked in the $J$ band data. H$_2$O features frequently seen in the spectra of L dwarfs of similar temperatures are not apparent in the data.
    \label{fig:nires}
\end{figure*}

In mid-infrared, the Wide-field Infrared Survey Explorer satellite (WISE, \citealp{Wright_WISE}) has observed the field containing ZTF J1239+8347 550 times in the \textit{W1} and \textit{W2} bands. We utilize the WISE/NEOWISE Coadder tool \citep{Masci_WISE} to stack all available WISE images by band. From this we retrieve an AB \textit{W1} magnitude of $21.5$ $\pm 1.0$ and do not detect the object in \textit{W2}. WISE images are taken at random orbital phases. The photometric measurement obtained from WISE thus represents an orbital average.

We obtained a NIR spectrum of ZTF J1239+8347 in JHK with the Keck Near-Infrared Echellete Spectrometer (NIRES, \citealp{Wilson_NIRES}) on 2025 September 4 over 3 hours, of which 1 hour provided useable data. The data, which have spectral resolution $R \sim 2700$, were reduced using a custom version of the IDL-based reduction package {\tt Spextool} \citep{Cushing_SPEX} modified for use with NIRES and using {\tt xtellcor} \citep{Vacca_Tellurics} to correct for telluric features in the spectrum using an A0 standard star observed close in airmass and time to the target. We cut to the range of valid detector response and remove outlier points from the combined spectrum by removing points that are more than $5 \sigma$ away from a 9 degree polynomial continuum.

The NIRES spectrum and best fit model are shown in figure \ref{fig:nires}. The accretion spectrum is apparent from the blue side, with additional structure noticeable in the J and H bands from the atmosphere of one or more BDs. We compare two model fits to the NIRES spectrum: a model consisting of only a blackbody fixed at $T = 8900$ K scaled by a free parameter for the accretion radius, $\Big(\frac{R_{\rm spot}}{d} \Big)^2$ where $d$ is the known distance to the system, and a mixture model of the same blackbody plus a single brown dwarf atmospheric model scaled by a second free parameter for the BD radius, $\Big(\frac{R_{\rm BD}}{d} \Big)^2$

\begin{equation} \label{eq:bb_bd_mixture_model}
    F_\mathrm{model}(\lambda) = \, \Big(\frac{R_{\rm spot}}{d} \Big)^2 F_\mathrm{BB}(\lambda, 8900) + \, \Big(\frac{R_{\rm BD}}{d} \Big)^2 F_\mathrm{BD}(\lambda, T_\mathrm{atmo})
\end{equation}

\noindent We utilize a grid of \texttt{BT-Settl} \citep{Allard_BT-Settl} cloudless BD atmospheres at temperatures between $T_\mathrm{atmo} = 1000 - 3500$ K with solar metallicity and $\log(g) = 4.5$. We use least squares to fit both scale parameters for each BD atmosphere model in our grid and retrieve a minimum $\chi^2$ for each model. We also compute the minimum $\chi^2$ of the single blackbody model without a brown dwarf atmosphere. The NIRES spectrum consists of $N = 5105$ data points We find a $\chi^2/\mathrm{dof} = 4.98$ \, for the single blackbody model with a single fit parameter, and that all BD atmosphere models between $T_\mathrm{atmo} = 1500 - 2600$ K with two fit parameters produce a $\chi^2/\mathrm{dof} \approx 4.63$. Over $5105$ data points, an improvement of $0.35$ dominates the Bayesian information criterion (BIC). From this we compute a BIC difference, $\Delta BIC \sim -1700$, indicating that a mixture model involving a brown dwarf atmosphere fits the observed NIRES spectrum significantly better than the single blackbody. We also compute a $\Delta BIC$ over the NIRES spectrum after it has been smoothed by a $5 \sigma$ Gaussian kernel and find that the $\Delta BIC$ does not change significantly.

In addition to this, we attempt to fit a similar mixture model consisting of a blackbody and two brown dwarf atmospheres scaled by the same parameter:
\begin{equation} \label{eq:bb_two_bd_mixture_model}
\begin{gathered}
F_\mathrm{model}(\lambda) = \, \Big(\frac{R_{\rm spot}}{d} \Big)^2 F_\mathrm{BB}(\lambda, 8900) \\
+ \, \Big(\frac{R_{\rm BD}}{d} \Big)^2 \Big( F_\mathrm{BD,1}(\lambda, T_\mathrm{eff1,atmo}) 
+ F_\mathrm{BD,2}(\lambda, T_\mathrm{eff2,atmo}) \Big)
\end{gathered}
\end{equation}
\noindent We find that a mixture model consisting of two BD atmospheres and a blackbody does not provide a statistically significant better fit to the observed spectrum than a single BD atmosphere and a blackbody. This is not unexpected given the low SNR of the NIRES spectra. The implications of this model fitting for the atmosphere of the secondary is discussed in section \ref{sec:results}.

The NIRES spectrum shows evidence of Na and K absorption features in $J$ band, typical of an L dwarf \citep{Cushing_2005}. L dwarfs at the derived temperature of the system typically also show H$_2$O features, as seen in the model in figure \ref{fig:nires} just blueward of the Na I doublet, which are not apparent in the observed spectrum.

\section{Results and System Properties} \label{sec:results}

Optical light curves of the system show steep positive excursions while optical spectra show a distinct absence of emission features. These phenomena are indicative of a direct impact configuration \citep[][T. Marsh, private communication]{Marsh_V407_DI, Marsh_Direct_Impact, Barros_DI, Wood_DI}. We confirm that aBDBs will be direct impact accretors by computing the minimum ballistic stream trajectory distance with equation 6 in \citet{Nelemans_WD_Synthesis}. We adopt orbital separations of two BDs of plausible masses between 10 -- 80 $\MJup$ at the period of ZTF J1239+8347. We find that for all possible BD mass configurations the minimum ballistic stream trajectory distance is less than the accretor radius and thus the system is physically consistent with direct impact accretion.

The RVs measured from the LRIS spectra represent the center of light of the absorption features from the hotspot. These absorption features can be subject to phase-dependent line profile changes which bias RV measurements \citep{Burdge_BW}. Additionally these RVs do not directly track the orbital motion of the accretor and require geometric and optical depth corrections dependent on the unknown spot latitude and inclination of the system \citep{Parsons_COL1, Parsons_COL2}. For these reasons we do not attempt to measure the orbital RV semi-amplitude as the scatter on the measured RV epochs significantly exceeds the formal uncertainties which indicates that phase-dependent line profile changes are included in the RV measurements. We note, however, that the scatter in individual RVs implies an RV semi-amplitude $K_{\rm COL} \lesssim 200$\,km\,s$^{-1}$.

While it is difficult to determine masses without an RV ratio of the system components, we estimate that the components of ZTF J1239+8347 each have masses between $60 - 80$ $\MJup$ based on BD isochrones \citep{Phillips_BD_Atmo_Models} with the accretor being more massive than the donor, which is consistent with the temperature range derived from the infrared spectra.

The shape of the apparent BD atmospheric spectrum only loosely constrains the temperature to $T_\mathrm{atmo} = 1500 - 2600$, however the best fit BD atmosphere scale parameter provides a fit for the radius. A real BD in the system must have a radius at least as large as the minimum radius of a BD ($\approx 0.95 \, \RJup$), and is likely slightly inflated relative to a typical BD at the system's age. We find the most plausible temperature to be $T_{\rm eff} \approx 1500$ K with a BD radius of $R_* = 1.20^{+0.15}_{-0.11} \, \RJup$, consistent with a typical $80 \, \MJup$ BD at ages $t_* \geq 1$ Gyr, or a moderately inflated BD of lower mass. The fit radius of the BD atmospheric component is consistent with an inflated BD, and none of the plausible temperatures provide radii consistent with two equal temperature components. This indicates that the dimmer component of the binary has $F_{\rm BD, 2} \lesssim 0.5 \times 10^{-18}$ erg/s/cm$^2$/\AA \, in J band and is therefore $T_{\rm eff} \lesssim 1200$ K. 
Assuming that the accretor, being the component that is heated from mass transfer, is the hotter of the components suggests the visible atmosphere belongs to the accretor.

In general, it is not guaranteed that the stream will strike the equator of the accretor. BD magnetic fields can be strong \citep{Reiners_BD_Magnetic_Fields}, and they or the Coriolis force could deflect the accretion stream to strike the accretor at some latitude. We attempt to constrain the radius, $R_\mathrm{spot}$, and colatitude, $\lambda$ of the hotspot by modeling the light curve with the python package \texttt{PHOEBE} \citep{Prsa_PHOEBE1, Prsa_PHOEBE2}. We fit synthetic light curves generated over a range of $R_\mathrm{spot}$ and $\lambda$ parameters to measure a joint likelihood for the parameters. We fix the inclination to three representative possible values, $i = 30, 45, 60 \degree$, and fix the relative intensity of the spot to the surface atmosphere. We let the spot longitude parameter and total intensity float, as these only affect the phase of the observation and the scale of the light curve. We do not account for tidal deformation from the donor, as this is a secondary effect on the spot radius. The spot contrast is fixed by assuming that the BD accretor photosphere contributes 0 flux in optical. We obtain a $\chi^2$ for a grid of $R_\mathrm{spot}$ and $\lambda$ for each $i$. We convert these values to likelihoods, $\mathcal{L}$, and normalize. We compute a cumulative distribution function (CDF) based on the $\mathcal{L}$ grid to compute a $1 \sigma$ contour.

The joint likelihood favors an equatorial spot for each inclination, with $\sim 45, 20, 20 \degree$ variance from the equator for $i = 30, 45, 60$ respectively. The spot radius is loosely constrained by the light curve shape. The best fit radii are $43, 66, 80$ respectively, but the spot could extend to over half the surface of the accretor depending on the inclination of the system and the asymmetry of the spot. A moderate or large spot radius is supported by the scale parameter and distance from equation \ref{eq:bb_bd_mixture_model}. For a BD of $T_{\rm eff} = 1500$ K the best fit hotspot flux scale parameter converts to a radius of $R_{\rm spot} = 74 \pm 8.4 \degree$. The radius of the spot is itself only weakly related to the radius of the accretion stream, and could be strongly asymmetrical, as strong convection and advection in BDs possibly causes diffusion around the area of contact of the accretion stream \citep{Mukherjee_Mixing}. A slight asymmetry in the system light curve (Figure~\ref{fig:photometry}) could be explained by a longitudinally extended hotspot. In addition, if the majority of the energy deposition occurs deep enough in the accretor's atmosphere, the spot will appear larger and dimmer on the surface. Hydrogen absorption features in the optical spectrum implies that the source of radiation is obscured by a substantial amount of hydrogen gas, i.e. that it is radiated inside the envelope.

Within the determined mass range, conventional cooling models would rule out old systems. The coolest and lowest mass plausible BD (e.g. $T_\mathrm{atmo} = 1500$, $M = 60 \MJup$) could be no older than 2 Gyrs, while at the highest temperature the system would be $\sim 100$ Myrs \citep{Phillips_BD_Atmo_Models}. Age estimates based on luminosity implicitly assume that these aBDB systems follow normal cooling tracks, which may not be the case. Accretion heating could potentially reverse cooling in the accretor, making conventional cooling not applicable.

\begingroup
\setlength{\tabcolsep}{6pt}
\renewcommand{\arraystretch}{0.92}
\tabletypesize{\footnotesize}

\begin{deluxetable}{lcc}
\tablecaption{System Properties \label{table:Observables}}
\vspace{-6pt} 
\tablewidth{0pt}
\tablehead{\vspace{-6pt}} 
\startdata
\multicolumn{3}{c}{\textbf{Astrometry} \tablenotemark{a}} \\
\hline
Right Ascension [J2016]  & $\alpha$ [deg] & 189.9798 \\
Declination [J2016]     & $\delta$ [deg]  & 83.7853 \\
Distance                & $d$ [pc]        & $339^{+42}_{-31}$ \\
Proper Motion (RA)      & $\mu_\alpha$ [mas yr$^{-1}$] & $-22.7 \pm 0.5$ \\
Proper Motion (Dec)     & $\mu_\delta$ [mas yr$^{-1}$] & $-19.8 \pm 0.4$ \\
\hline
\multicolumn{3}{c}{\textbf{Photometry}} \\
\hline
Mean HiPERCAM g Magnitude & $m_g$ [AB] &  $20.3 \pm 0.7$\\
Mean HiPERCAM i Magnitude & $m_i$ [AB] &  $21.0 \pm 0.5$\\
Mean WIRC J Magnitude & $m_J$ [AB] &  $20.3 \pm 0.8$\\
Mean WIRC K Magnitude & $m_K$ [AB] &  $21.2 \pm 1.2$\\
Mean WISE 1 Magnitude & $m_{W1}$ [AB] & $21.5 \pm 1.1$\\
\hline
\multicolumn{3}{c}{\textbf{Physical Properties}} \\
\hline
Donor Mass   & $M_d$ [$M_{\rm Jup}$] &  $[60, 80]$\\
Accretor Mass & $M_a$ [$M_{\rm Jup}$] &  $[60, 80]$\\
Accretor Radius & $R_a$ [$R_{\rm Jup}$] & $1.20^{+0.15}_{-0.11}$\\
Donor Radius & $R_d$ [$R_{\rm Jup}$] & $[0.9, 1.4]$\\
Accretor Atmospheric Temperature & $T_{\rm a}$ [K] & $\sim 1500$\\
Donor Atmospheric Temperature & $T_{\rm d}$ [K] & $\lesssim 1200$\\
Hotspot Effective Temperature & $T_{\rm spot}$ [K] & $8904 \pm 54$\\ 
Age (conventional)\tablenotemark{b} & $t_*$ [Gyr] & $[0.1, 2]$\\
Spot Radius & $R_{\rm spot}$ [$R_{\rm Jup}$] & $0.60^{+0.07}_{-0.06}$\\
\enddata
\tablenotetext{a}{\citep{GaiaDR3}}
\tablenotetext{b}{Conventional cooling models may be unreliable under the affects of accretion heating.}
\end{deluxetable}
\endgroup

\section{Discussion} \label{sec:discussion}
\begin{figure}[t!]
    \centering
    \includegraphics[width=1\linewidth]{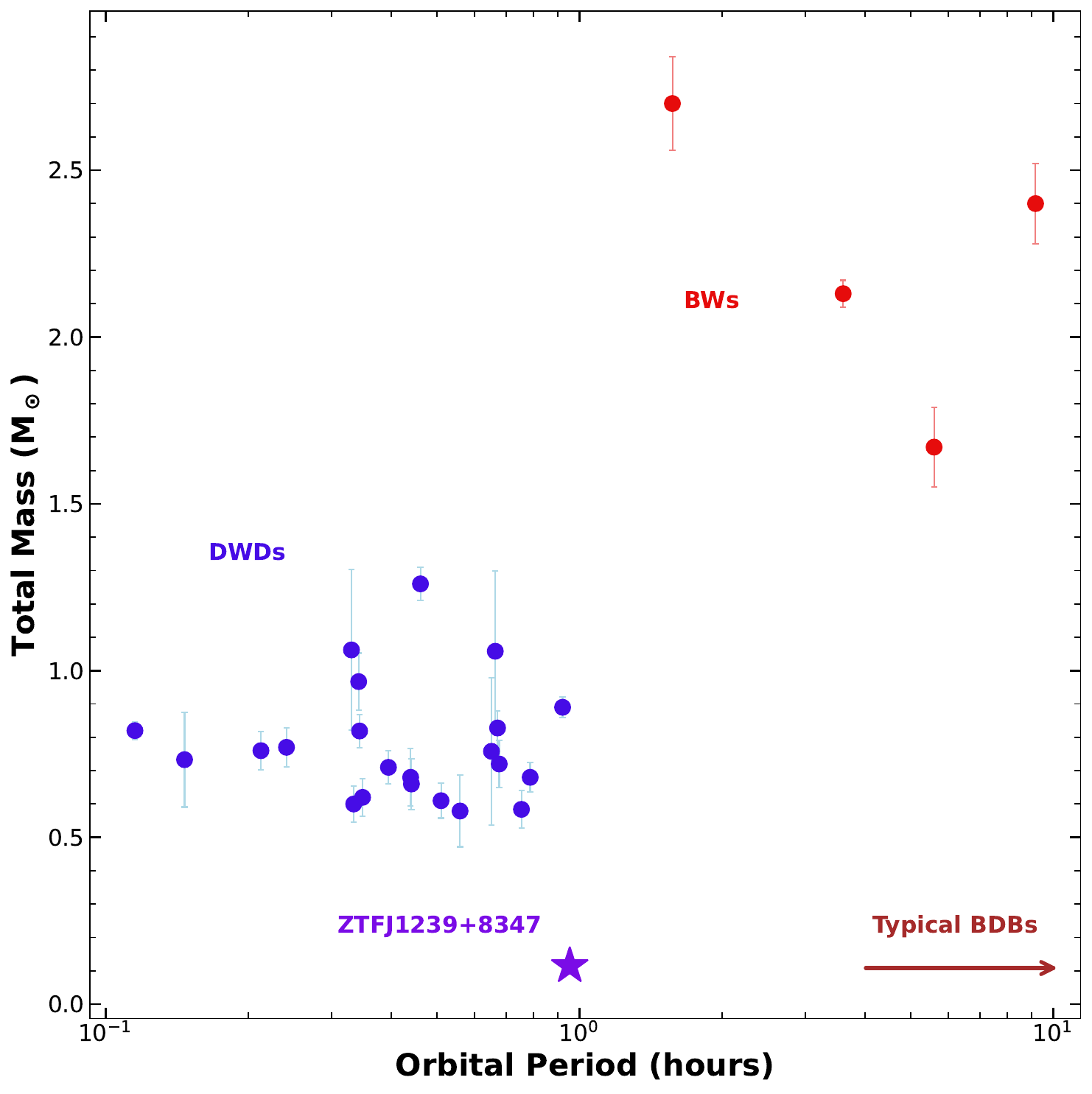}
    \caption{Comparison of ZTF J1239+8347 to double white dwarfs (DWDs) and black widow neutron star -- substellar object binaries (BWs). We note that typical brown dwarf binaries (BDBs) exist at much longer periods than ZTF J1239+8347.}
    \label{fig:comparison}
\end{figure}

ZTF J1239+8347 provides a potentially valuable probe of the dynamics of stable mass transfer at the lowest detectable mass scales. Figure \ref{fig:comparison} shows ZTF J1239+8347 in a period-mass comparison to other compact binaries \citep{Munday_DWD, Kennedy_BW, van_Kerkwijk_BW, Romani_BW1, Romani_BW2} with periods below 10 hours given that ZTF J1239+8347 is an aBDB. Most known BDBs have periods greater than $10^{5}$ hours. The shortest to date known non-interacting BDB is LP 413-43AB \citep{Hsu_short_BDB} at 17 hours. 

Optical spectra of the system exhibit H-$\alpha$ emission near minimum brightness (Figure~\ref{fig:lris}). This emission feature is unlikely to arise from the hotspot, which is primarily out of view at this phase. It is possible that this feature is caused by reprocessing in the accretor atmosphere which is significantly dimmer than the hotspot, or by reflection off the surface of the donor.

We simulate AML in aBDBs from gravitational radiation \citep{landau1975classical} and several magnetic braking prescriptions \citep{Sills_MB, Belloni_MB} and determine that magnetic braking is the likely dominant angular momentum loss (AML) mechanism in aBDB systems once they are in $\sim 1$ hr orbits. The efficiency of magnetic braking in binary AML involving low mass stars and substellar objects is still a matter of debate in literature, particularly as it relates to cataclysmic variable systems \citep{El-Badry_MB_Saturates, Ortuzar_MB_CVs, Schreiber_WDMF}. aBDBs then provide a valuable probe of the efficiency of magnetic braking in low mass systems. 

Future observations of the system with the James Webb Space Telescope (JWST) could constrain the temperature of the accretor atmosphere better, and could detect the atmosphere of the donor system. Such observations will be necessary to directly measure the mass ratio of the system, as it is not feasible for ground based near-infrared spectra to be phase resolved for systems at this brightness. More detailed study of the hotspot in particular would provide constraints on the mass transfer rate of the system, as well as test mixing in the accretor by measuring relative abundances of common BD atmospheric molecules on the day and night sides of the accretor.

ZTF J1239+8347 is near the detectability limit of ZTF. Based on the object's short distance from Earth, we expect the Vera Rubin observatory \citep{Ivezic_Rubin} to detect dozens more of these objects.

\subsection{Compact Binary Confounding} \label{subsec:compact_binaries}
All non-BD, non-compact object accretion engines would have enough atmospheric flux of their own to appear in the optical and infrared spectra of the system, so they can be ruled out, but could one of the components of ZTF J1239+8347 be a compact object? The data rules these out as well. ZTF J1239+8347 photometry appears similar to some known LMXB black widow systems \citep{Bobakov_Optical_BWs}, and a hypothetical neutron star ablating a substellar mass companion could produce the photometric variability observed in the system, but the \textit{Gaia} distance combined with the non-detection of X-Ray flux with \textit{Swift} limits the X-Ray luminosity to be $L_X \lesssim 1 \times 10^{30}$ erg/s, which is both an order of magnitude below what would be expected from a BW pulsar \citep{Bogdanov_Pulsar_X_Rays, Bahramian_LMXB}, and insufficient to produce the observed $T_\mathrm{acc} = 8900$ K on the dayside of a BD companion at the observed period. Further, the measured epoch RVs imply a $K_{\rm COL} \lesssim 200$\,km\,s$^{-1}$. Given the existence of at least one BD in the system, this semi-amplitude is inconsistent with a NS host, requiring an inclination $i \lesssim 5 \degree$. The nearest radio-loud confirmed black widow binary is $\sim 500$ pcs away \citep{Koljonen_BW_Survey}, and it is unlikely there is a closer undiscovered population. Additionally, ZTF J1239+8347 possesses a mild proper motion $\mu = 30$ mas yr$^{-1}$, at odds with the expected natal kick from a NS formation.

A Cataclysmic Variable system with a white dwarf accreting material from a brown dwarf is also impossible for several reasons. A CV system would see the 8900 K blackbody coming from accretion onto a hypothetical WD. Given the strong distance constraint, the effective radius of the emitting area of the blackbody is constrained to $R_{\rm BB} = 0.60^{+0.07}_{-0.06} \, \RJup$. This large of an emitting area is inconsistent with a canonical WD. In addition, the optical spectra lack the cyclotron humps characteristic of WD polar accretors \citep{Cropper_Polars, Van_Roestel_Polars}. Additionally in this configuration it is impossible for the hot spot to be on an irradiated BD, as the irradiating WD would be visible in the optical spectrum at all times.

These similarities to compact object cases do, however, present a difficulty in the search for further aBDB systems. \textit{Gaia} photometry places ZTF J1239+8347 in the regime of CV systems, which would make similar systems easy to miss.

\section{Acknowledgments}\label{sec:acknowledgements}
We thank the anonymous reviewer for their thorough comments. We thank Cheyanne Shariat, Shri Kulkarni, Jim Fuller, Soumyadeep Bhattacharjee, Kareem El-Badry, and April Luce for their insights and comments. We also acknowledge the initial work of Thomas Marsh in consideration that mass transfer between brown dwarfs could appear similar to black widow binaries in optical photometry. We also wish to acknowledge Matthew Graham and the ZVAR collaboration for their contribution to the detection of periodic variable stars. Some of the data presented herein were obtained at Keck Observatory, which is a private 501(c)3 non-profit organization operated as a scientific partnership among the California Institute of Technology, the University of California, and the National Aeronautics and Space Administration. The Observatory was made possible by the generous financial support of the W. M. Keck Foundation. The authors wish to recognize and acknowledge the very significant cultural role and reverence that the summit of Maunakea has always had within the Native Hawaiian community. 

This publication makes use of data products from the Wide-field Infrared Survey Explorer, which is a joint project of the University of California, Los Angeles, and the Jet Propulsion Laboratory/California Institute of Technology, funded by the National Aeronautics and Space Administration. This dataset or service is made available by the Infrared Science Archive (IRSA) at IPAC, which is operated by the California Institute of Technology under contract with the National Aeronautics and Space Administration. 

This work was partially supported by funding from the NASA LISA Preparatory Science grants 80NSSC24K0361 and 80NSSC21K1723, as well as the NASA FINESST Graduate Student Research grant 80NSSC23K1434. J.W.X is grateful for support from the Heising-Simons Foundation 51 Pegasi b Fellowship (grant \#2025-5887).

\bibliographystyle{aasjournal}
\bibliography{refs.bib}
\label{lastpage}
\end{document}